body of paper
--------------------------------------------------------------
\documentstyle[12pt]{article}
\font\bfmis=cmbx12

\newcommand{\be}{\begin{equation}}
\newcommand{\ee}{\end{equation}}
\newcommand{\bd}{\begin{displaymath}}
\newcommand{\ed}{\end{displaymath}}
\begin{document}
\begin{titlepage}
\vspace{3cm}
\centerline{\bfmis GALILEAN INVARIANCE}
\centerline{\bfmis IN 2+1 DIMENSIONS}
\vspace{2cm}
\centerline{Yves Brihaye}
\centerline{Dept. of Math. Phys.}
\centerline{University of Mons}
\centerline{Av. Maistriau, 7000 Mons, Belgium}
\vspace{1cm}
\centerline{Cezary Gonera\footnote{supported by the grant
Nr 458 of University of \L\'od\'z}\footnote{Permanent
address: Dept. of Theor. Phys., Univ. of
\L\'od\'z, Pomorska
149/153, 90-236 \L\'od\'z, Poland}}
\centerline{Dept. of Physics}
\centerline{U.J.A}
\centerline{Antwerpen, Belgium}
\vspace{1cm}
\centerline{Stefan Giller*, Piotr Kosi\'nski*}
\centerline{Dept. of Theoretical Physics}
\centerline{University of \L\'od\'z}
\centerline{Pomorska 149/153, 90-236 \L\'od\'z, Poland}
\vspace{2cm}
\noindent
\underline{ABSTRACT:}\newline The Galilean invariance in
threedimensional space-time is considered. It appears that
the Galilei group in 2+1 dimensions posses a
three-parameter family of projective representations. Their
physical interpretation is discussed in some detail.
\end{titlepage}

It is well known [1] that, contrary to the case of Poincare
group, the Galilei group posses a family of nontrivial
projective representations. Moreover, it has been shown
[2], [3] that precisely these projective representations
are physically meaningful. In particular, with the wave
functions transforming according to the true
representations of Galilei group one can construct no
localized states [2] and no reasonable position operator
[3].

The projective representations of $3+1$-dimensional Galilei
group were studied in some detail by Levy-Leblond [4]. The
family of nonequivalent projective representations is
parametrized by one real parameter which is to be
identified with the particle mass (the negative mass case
calls for some reinterpretation-see Ref.~[4]).

Recently,
some attention has been paid to the representations of
Galilei group in $2+1$ dimensions [5], [6]. It appears
that, mainly due to the extremely simple structure of the
rotation group in two dimensions, the $2+1$-dimensional
Galilei group admits a three-parameter family of
nonequivalent projective representations (for a more
precise statement see [6]).

In this note we discuss in more detail the possible
physical meaning of these projective representations. We
conclude that some of them should be simply rejected while
the others give rise to quite interesting phenomena. Let us
start with a brief description of $2+1$-dimensional Galilei
group. It is defined as the group of the space-time
transformations of the form
\be
\left\{
\begin{array}{l}
\vec x'=\overrightarrow{W(\Theta)x}+\vec vt+\vec u\\
t'=t+\tau
\end{array}
\right.
\ee
where
\be
W(\Theta)=\left(
\begin{array}{cc}
\cos\Theta & \sin\Theta\\
-\sin\Theta &\cos\Theta
\end{array}
\right)
\ee
is a rotation. The universal covering group is parametrized
by six real parameters $(\Theta,\tau,\vec v,\vec u)$
subject to the following composition law
\begin{eqnarray}
\lefteqn{(\Theta',\tau',\vec v',\vec u')*(\Theta,\tau,\vec
v,\vec u)=}\nonumber\\
& &
(\Theta'+\Theta,\tau'+\tau,\overrightarrow{W(\Theta')v}+\vec
v',\overrightarrow{W(\Theta')u}+\tau\vec v'+\vec
u')
\end{eqnarray}

The following matrix realisation is often useful\newpage
\begin{eqnarray}
\lefteqn{\left(
\begin{array}{ccc}
W(\Theta) & \vec v & \vec u \\
0 & 1 & \tau \\
0 & 0 & 1
\end{array}
\right)=}\nonumber \\
& & \left(
\begin{array}{ccc}
1 & 0 & 0 \\
0 & 1 & \tau \\
0 & 0 & 1
\end{array}
\right)
\left(
\begin{array}{ccc}
1 & 0 & \vec u \\
0 & 1 & 0 \\
0 & 0 & 1
\end{array}
\right)
\left(
\begin{array}{ccc}
1 & \vec v & 0 \\
0 & 1 & 0 \\
0 & 0 & 1
\end{array}
\right)
\left(
\begin{array}{ccc}
W(\Theta) & 0 & 0 \\
0 & 1 & 0 \\
0 & 0 & 1
\end{array}
\right)
\end{eqnarray}
Accordingly, there exists the following \underline{global}
exponential parametrization
\be
g=e^{-i\tau H} e^{i\vec u\vec P}e^{i\vec v\vec K}e^{i\Theta
J}
\ee
The corresponding Lie algebra reads
\bd
[J,P_i]=i\varepsilon_{ij}P_j
\ed
\be
[J,K_i]=i\varepsilon_{ij}K_j
\ee
\bd
[K_i,H]=iP_i
\ed
all remaining commutators being vanishing.

In order to find projective representations of Galilei group
one has to study the central extensions of the above
algebra. This is rather easy. One adds to the right-hand
sides of all commutation rules a central element {\bf1}
with arbitrary coefficients. After (i) making an
appropriate redefinitions of P's and K's and (ii) using
Jacobi identities one arrives at the following
three-parameter family of central extensions
\bd
[P_i,K_j]=-im\delta_{ij}{\bf1}
\ed
\bd
[K_i,K_j]=ik\varepsilon_{ij}{\bf1}
\ed
\be
[J,H]=ig{\bf1}
\ee
\bd
[K_i,H]=iP_i
\ed
\bd
[J,P_i]=i\varepsilon_{ij}P_j
\ed
\bd
[J,K_i]=i\varepsilon_{ij}K_j{\rm,}
\ed
remaining commutators being vanishing. Let us note some
properties of the above algebra:\newline
(i) if $g=0$ one can obtain the above structure from the
contraction of $2+1$-dimensional Poincare algebra. The
latter reads
\bd
[J,P_i]=i\varepsilon_{ij}P_j
\ed
\bd
[J,L_i]=i\varepsilon_{ij}L_j
\ed
\be
[J,P_0]=0
\ee
\bd
[L_i,L_j]=-i\varepsilon_{ij}J
\ed
\bd
[L_i,P_j]=i\delta_{ij}P_0
\ed
\bd
[L_i,P_0]=iP_i
\ed
Now we add one central element {\bf1} to obtain the direct
sum of Poincare algebra and onedimensional one and redefine
\bd
P_0\rightarrow mc{\bf1}+H/c
\ed
\be
J\rightarrow -kc^2{\bf1}+J
\ee
\bd
L_i\rightarrow cK_i
\ed
Letting the contraction parameter $c\rightarrow\infty$ we
get (7). The case $g\neq0$ cannot be obtained by
contraction from Poincare algebra.\newline
(ii) It is extremely important to observe that, for
$m\neq0$, one can get rid of the parameter $k$ by
redefining
\be
K_i\rightarrow K_i+\frac{k}{2m}\varepsilon_{ij}P_j{\rm;}
\ee
all remaining commutators are unaffected by this
transformation. This remark plays an important role in what
follows. \newline
(iii) From the Casimir operators for Poincare group
\bd
\tilde C_1=P_0^2-\vec P^2
\ed
\bd
\tilde C_2=P_0J+\varepsilon_{ij}P_iL_j
\ed
one obtains by contraction procedure the Casimir operators
for the case $g=0$:
\bd
C_1=H-\vec P^2/2m \mbox{   (internal energy) }
\ed
\be
C_2=J-\frac{1}{m} \vec K\times\vec P-\frac{kH}{m} \mbox{
   (spin) }
\ee
The case $g\neq0$ is slightly more involved. Assume first
that $m\neq0$. We can use eq.(10) to get rid of $k$. Now,
let $C(\vec K,\vec P,H,J)$ be a Casimir operator. Using (7)
we get (with $C_{1,2}$ given by eq.(11) with $k=0$)
\bd
e^{i\lambda C_1}C(\vec K,\vec P,H,J)e^{-i\lambda C_1}=
C(\vec K,\vec P,H,J+\lambda g{\bf1})
\ed
\be
e^{i\lambda C_2}C(\vec K,\vec P,H,J)e^{-i\lambda C_2}=
C(\vec K,\vec P,H-\lambda g{\bf1},J)
\ee
Therefore $C$ cannot depend on $H$ and $J$. Applying the
same reasoning with $C_i$ replaced by $K_i$ or $P_i$ we see
that $C$ must be a constant.
Assume now that $m=0$; then ${\vec P}^{\,2}$ is the Casimir
operator. Moreover, for $k=0$ (remember that for $m=0$ we
cannot put $k=0$ by redefining of boosts), $\vec
K\times\vec P$ is also the Casimir operator. It is now easy
to find the composition law for projective representations.
Keeping in mind that one should put ${\bf1}\rightarrow1$
within representation and using
\be
U(g)=e^{-i\tau H}e^{i\vec u\cdot\vec P}e^{i\vec v\cdot\vec
K}e^{i\Theta J}
\ee
together with the commutation rules (7) we get
\bd
U(g')U(g)=\omega(g',g)U(g'g)
\ed
\be
\omega(g',g)=e^{ig\tau\Theta'}e^{\frac{-i}{2}m\tau\vec{v'}^2}
e^{-im\vec v'\cdot\overrightarrow{\scriptstyle W(\Theta')u}}
e^{\frac{-i}{2}k(\vec
v'\times\overrightarrow{\scriptstyle W(\Theta')v})}
\ee
Let us now discuss the physical interpretation of the
representations under consideration. First of all one can
argue that the case $g\neq0$ is unphysical. Indeed, the
Casimir operators either do not exist (if $m\neq0$) or do
not depend on $H$ (if $m=0$). Therefore, there exists no
counterpart of Schr\"odinger equation in $\vec x$-space;
there will be no dynamics. This conclusion is supported by
the form of ireducible representations for this case [6].
They either contain any (square integrable) function
$f(\vec p,\varepsilon)$ (if $m\neq0$) or are concentrated
on submanifold $\vec p=const$, again with arbitrary
$\varepsilon$-dependence. Translated to $t-\vec x$-space it
means that the time behaviour of wave functions can be
arbitrary. Let us now concentrate on $g=0$ case. It is easy
to find the irreducible representations using standard
methods. Assume that $m\neq0$. The extended Galilei group
acts transitively on the paraboloid $\varepsilon-\frac{\vec
p^2}{2m}=v$, $v$ being internal energy. As in the
$3+1$-dimensional case one can argue that putting $v=0$
does not restrict generality [4]. We choose $\vec p=0$ as a
standard vector and define an arbitrary eigenvector $|\vec
p,s\rangle$ (s to be specified below) by
\be
|\vec p,s\rangle=B(\vec p)|\vec 0,s\rangle
\ee
where
\be
B(\vec p)\equiv e^{\frac{i}{m}\vec p\cdot\vec K}
\ee
is the standard boost. Noting that
\be
U^+(g)\vec PU(g)=\overrightarrow{W(\Theta)P}+m\vec v{\bf1}
\ee
and using the substitution rule ${\bf1}\rightarrow1$ we
write
\be
U(g)|\vec p,s\rangle=B({\vec p}^{\,\prime})[B^+({\vec
p}^{\,\prime}) U(g)B(\vec
p)]|\vec 0,s\rangle
\ee
where ${\vec p}^{\,\prime}=\overrightarrow{W(\Theta)p}+m\vec v$. The
expression in square bracket is an element of the little
group of $\vec p=\vec 0$. But this group is generated by
$J$ and ${\bf1}(=1)$ and is therefore characterized by one
number $s$ (spin). Using eq.(7) we arrive at the following
explicit form of the representation:
\begin{eqnarray}
\lefteqn{U((\Theta,\tau,\vec v,\vec u))|\vec
p,s\rangle=}\nonumber\\
& & e^{i(-\frac{\tau}{2}m\vec v^2-\frac{\tau}{2m}\vec
p^2+m\vec u\vec v+(\vec u-\tau\vec
v)\overrightarrow{\scriptstyle W(\Theta)p}-\frac{k}{2m}(\vec
v\times\overrightarrow{\scriptstyle W(\Theta)p})+\Theta
s)}\nonumber\\
& & |\overrightarrow{W(\Theta)p} +m\vec v,s\rangle
\end{eqnarray}
The states $|\vec p,s\rangle$ are normalized in the
invariant manner by $\langle\vec p,s|{\vec
p}^{\,\prime},s\rangle=\delta^{(2)}(\vec p-{\vec p}^{\,\prime})$. The wave
function $f(\vec p,s)$ is defined by
\bd
|f\rangle=\int d^2\vec pf(\vec p,s)|\vec p,s\rangle
\ed
\be
f(\vec p,s)=\langle\vec p,s|f\rangle
\ee
The action of Galilei group on wave functions can be read
off from eqs.(19) and (20). We get:\newline
-translations:
\setcounter{equation}{21}
$$
f(\vec p,s)\rightarrow e^{-i\tau\frac{\vec p^2}{2m}}
e^{i\vec u\vec p}f(\vec p,s) \eqno (\theequation a)
$$
-rotations
$$
f(\vec p,s)\rightarrow e^{i\Theta
s}f(\overrightarrow{W(-\Theta)p},s)\eqno(\theequation b)
$$
-boosts
$$
f(\vec p,s)\rightarrow e^{-\frac{ik}{2m}(\vec v\times\vec
p)} f(\vec p-m\vec v,s)\eqno(\theequation c)
$$
Correspondingly, the generators read
\bd
\vec P=\vec p{\rm,}\quad H=\frac{{\vec p}^{\,2}}{2m}
\ed
\be
J=-i(\vec p\times\vec\nabla_p)+s
\ee
\bd
K_i=im\frac{\partial}{\partial
p_i}-\frac{k}{2m}\varepsilon_{ij}p_j
\ed
As we have noted above, one can always redefine $K_i$'s
in such a way that $k$ disappears from the algebra.
However, such a transformation provides rather
a redefinition of physical observables than a canonical
transformation. Therefore, we will also discuss the case
$k\neq0$. Due to the commutation rules $[K_i,H]=iP_i$,
$[K_i,P_j]=im\delta_{ij}\cdot{\bf1}$ it is tempting to
define the position operator as $X_i=\frac{1}{m}K_i$.
However, this is wrong. To see this let is invoke the
natural requirements the position operator should obey [3]:
(i) its expectation value should change by $\vec a$ when we
go from any state to the state obtained by translation
through $\vec a$, (ii) it should transform like a vector
under rotation and (iii) it should be unchanged by boosts.
These conditions are met by the following choice
\be
X_i=\frac{1}{m}K_i+\frac{k}{m^2}\varepsilon_{ij}P_j
\ee
It appears now that the position operators do not commute,
their commutator being
\be
[X_i,X_j]=-\frac{ik}{m^2}\varepsilon_{ij}
\ee
Such a situation is not possible in three space dimensions
due to the lack of invariant second order antisymmetric
tensor.

Eq.(24) has far reaching consequences. It is obvious that
the momentum wave function preserves its probability
interpretation. Let us, however, consider the
Fourier-transformed wave function
\be
\tilde f_s(\vec x,t)=\int d^3\vec pf(\vec p,s)e^{i(\vec
p\cdot\vec x-\frac{\vec p^2}{2m}t)}
\ee
It obeys standard free Schr\"odinger equation. The action
of $X_i$ operators is easily found to be
\be
(X_i\tilde f_s)(\vec
x,t)=(x_i-\frac{ik}{2m^2}\varepsilon_{ij}\frac{\partial}{\partial
x_j})\tilde f_s(\vec x,t)
\ee
However, by obvious reasons $\tilde f_s(\vec x,t)$ lacks
its standard probability interpretation. Instead, the
following uncertainty principle holds
\be
\bigtriangleup X_1\cdot\bigtriangleup
X_2\geq\frac{|k|}{2m^2}
\ee
It is not difficult to find the wave functions saturating
this inequality. They read
\be
f(\vec p)=F(\gamma p_1-ip_2)e^{[u+\frac{k}{2m^2}(\gamma
p_1-ip_2)]p_1}
\ee
here $\gamma\in{\bf R}$, $u\in{\bf C}$  and
$F$ is arbitrary function chosen in such a way that $f(\vec
p)$ is normalizable.

In the classical limit one finds the hamiltonian theory
defined by the following Poisson bracket
\be
\{F,G\}=\frac{\partial F}{\partial x_i}\frac{\partial
G}{\partial p_i}-\frac{\partial F}{\partial
p_i}\frac{\partial G}{\partial
x_i}-\frac{k}{m^2}\varepsilon_{ij}\frac{\partial
F}{\partial x_i}\frac{\partial G}{\partial x_j}
\ee
summation over repeated indices being understood.

The conserved generators (their action being defined by the
above Poisson bracket) read
\bd
\vec P=\vec p{\rm,}\qquad H=\frac{{\vec p}^{\,2}}{2m}
\ed
\be
J=\vec x\times\vec p-\frac{k{\vec p}^{\,2}}{2m^2}{\rm, }\qquad
K_i=m(x_i-\frac{k}{m^2}\varepsilon_{ij}p_j)-p_it
\ee

The second terms in expressions for $J$ and $\vec K$ are
necessary in order to provide a proper transformation law
for $\vec x$.

Now the important point is to note that due to the remark
(ii) made above the classical and quantum theories defined
for $k\neq0$ can be rephrased in terms of standard ones
using the following substitution rules
\bd
\vec p=\vec p_s
\ed
\be
x_i=x_{si}+\frac{k}{2m^2}\varepsilon_{ij}p_{sj}
\ee
\bd
K_i=K_{si}-\frac{k}{2m}\varepsilon_{ij}p_{sj}
\ed
where the subscript ``s'' refers to standard theory. As we
have noticed above this does not imply that these theories
are equivalent from the physical point of view due to
different interpretation of basic observables.

As an example let us consider twodimensional harmonic
oscillator (actually, this system is \underline{not}
Galilei-invariant but it may be replaced easily by two
particles coupled to each other by harmonic force)
\be
H=\frac{{\vec p}^{\,2}}{2m}+\frac{m\omega^2{\vec x}^{\,2}}{2}
\ee
$H$ commutes with angular momentum $J$ (eq.(30)) so we can
look for common eigenvectors
\be
H|n,l\rangle=E_n|n,l\rangle{\rm,}\quad
J|n,l\rangle=l|n,l\rangle
\ee
However, we can consider the equivalent standard theory.
Under the substitution (31) I takes its standard form,
$J=J_s$ while the hamiltonian reads
\bd
H=\frac{\vec p^{\,2}_s}{2m_s}+\frac{m_s\omega^2_s}{2}\vec
x^{\,2}_s+ \gamma J_s
\ed
\be
\frac{1}{m_s}=\frac{1}{m}(1+\frac{k^2\omega^2}{4m^2})
\ee
\bd
\frac{\omega^2_s}{\omega^2}=(1+\frac{k^2\omega^2}{4m^2})
\ed
\bd
\gamma=\frac{k\omega^2}{2m}
\ed
This allows us to find the spectrum of $H$ by considering
the spectrum of ``standard'' harmonic oscillator.

Finally, let us consider the case $m=0$. The $k\neq0$ case
cannot be now obtained from $k=0$ one by a simple
redefinition. However, both for $k=0$ and $k\neq0$, ${\vec
P}^{\,2}$ is the Casimir operator. The irreducible
representations are therefore constrained to live on circle
(or point) $\vec p^2=const$. Consequently, they are not
localisable on ${\bf R}^2$. Moreover, there is again no
constraint on energy (relating it to other ``observables'')
which leads to arbitrary time behaviour of space-time wave
functions.
\newpage
\vspace{4cm}
{\bfmis References:}
\frenchspacing

[1] V. Bargmann, Ann. Phys. {\bf59}, 1 (1954)

[2] E. In\"onu, E. P. Wigner, Nuovo Cimento {\bf9}, 705
(1957)

[3] M. Hammermesh, Ann. Phys. {\bf9}, 518 (1960)

[4] J-M. Levy-Leblond, J. Math. Phys. {\bf4}, 776 (1963)

[5] J. Mund, R. Schrader, preprint Univ. Berlin, SFB288,
no. 74 (1992)

[6] D. R. Grigore, preprint IFA-FT-391-1993

\end{document}